\documentclass[aps,twocolumn,showpacs,superscriptaddress]{revtex4-1}
\usepackage[T1]{fontenc}
\usepackage{amsfonts,amsmath,amssymb,amsbsy}
\usepackage{graphicx,color}
\usepackage{times}
\usepackage[colorlinks=true, linkcolor=blue, citecolor=red, urlcolor=magenta]{hyperref}

\let\mathbf=\boldsymbol
\def\section#1{\medskip\noindent\textbf{#1}\par}
\makeatletter
\renewcommand{\fnum@figure}{\figurename~\textbf{\thefigure}}
\makeatother
\renewcommand{\figurename}{\textbf{Figure}}
\def\blue#1{\textcolor{blue}{#1}}
\def\blue#1{\textcolor{black}{#1}}
\begin{document}

\title{{\Large Spin-Cherenkov effect in a magnetic nanostrip with interfacial Dzyaloshinskii-Moriya interaction}}

\author{Jing Xia}
\affiliation{School of Electronics Science and Engineering, Nanjing University, Nanjing 210093, China}
\affiliation{Department of Physics, University of Hong Kong, Hong Kong, China}
\author{Xichao Zhang}
\affiliation{School of Electronics Science and Engineering, Nanjing University, Nanjing 210093, China}
\affiliation{Department of Physics, University of Hong Kong, Hong Kong, China}
\author{Ming Yan}
\affiliation{Department of Physics, Shanghai University, Shanghai 200444, China}
\author{Weisheng Zhao}
\email[]{weisheng.zhao@buaa.edu.cn}
\affiliation{Fert Beijing Institute, Beihang University, Beijing 100191, China}
\affiliation{School of Electronic and Information Engineering, Beihang University, Beijing 100191, China}
\author{Yan Zhou}
\email[]{yanzhou@hku.hk}
\affiliation{School of Electronics Science and Engineering, Nanjing University, Nanjing 210093, China}
\affiliation{Department of Physics, University of Hong Kong, Hong Kong, China}
\affiliation{Center of Theoretical and Computational Physics, University of Hong Kong, Hong Kong, China}

\begin{abstract}\bf\noindent
Spin-Cherenkov effect enables strong excitations of spin waves (SWs) with nonlinear wave dispersions. The Dzyaloshinskii-Moriya interaction (DMI) results in anisotropy and nonreciprocity of SWs propagation. In this work, we study the effect of the interfacial DMI on SW Cherenkov excitations in permalloy thin-film strips within the framework of micromagnetism. By performing micromagnetic simulations, it is shown that coherent SWs are excited when the velocity of a moving magnetic source exceeds the propagation velocity of the SWs. Moreover, the threshold velocity of the moving magnetic source with finite DMI can be reduced compared to the case of zero DMI. It thereby provides a promising route towards efficient SW generation and propagation, with potential applications in spintronic and magnonic devices.
\end{abstract}

\date{\today}

\maketitle


\noindent
The Cherenkov radiation of light occurs when a charged particle moves faster than the light speed within a medium~\cite{1}. This effect is named after Pavel Alekseyevich Cherenkov, who studied this phenomenon experimentally. Cherenkov radiation is used frequently in particle identification detectors in particle physics~\cite{2}. The Cherenkov effect is analogous to the sonic boom produced by shock waves propagating away from an aircraft, if its speed exceeds the sound velocity. Similar to the Doppler effect observed in different physical system~\cite{3,4,5,6,7}, the Cherenkov effect belongs to one of the fundamental phenomena induced by the radiation of moving sources. The Cherenkov-like effect of spin waves (SWs) has been theoretically studied in ferromagnets, which can be used to excite coherent SWs without the necessity of external alternating magnetic field or current~\cite{8,9}. Recently, the study of the influence of the antisymmetric exchange interaction, namely the Dzyaloshinskii-Moriya interaction (DMI), on magnetic excitations such as domain walls and vortex is one of the hottest topics in nanomagnetism and spintronics~\cite{10,11,12,13,14,15,16,17,18,19,20}. DMI is an antisymmetric interaction induced by spin-orbital coupling due to broken inversion symmetry in lattices or at the interface of magnetic films~\cite{12}, which has been measured for both magnetic interfaces~\cite{21,22} and bulk materials~\cite{23,24}. DMI facilitates the creation of topologically protected spin textures in chiral magnetic materials, i.e. magnetic skyrmions, which are favorable information carriers in the next-generation data storage and spin logic devices. On the other hand, theoretical~\cite{13,14,16} and experimental~\cite{10,15,25} studies have demonstrated that DMI leads to an asymmetrical spin-wave dispersion. DMI has also been measured in a wide range of materials including permalloy~\cite{22,26,27}. Specifically, the asymmetry in the formation of a vortex state in a permalloy nanodisk has been studied by micromagnetic simulations with interfacial DMI~\cite{22}.

\begin{figure}[t]
\centerline{\includegraphics[width=0.50\textwidth]{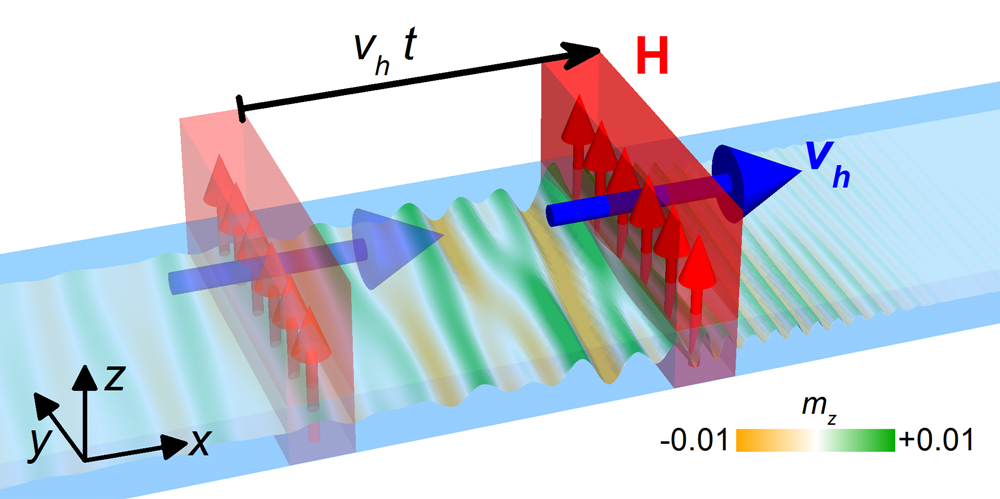}}
\caption{\textbf{Schematics of the micromagnetically modeled system.}
An external rectangular shape field pulse (\textbf{H}), $12$ nm wide in the $x$-direction and $100$ nm long in the $y$-direction, is applied to a $12$-$\mu$m-long, $100$-nm-wide, and $10$-nm-thick magnetic strip with a magnitude of $10$ mT in the $z$-direction and constant speed of $v_{h}$. The color scale represents the out-of-plane component of the magnetization $m_{z}$, which has been used throughout this paper.
}
\label{FIG1}
\end{figure}

In this paper, the influence of interfacial DMI on the Spin-Cherenkov effect (SCE) in permalloy strip are studied by micromagnetic simulations. Pictorial illustrations of the setup is shown in Fig.~\ref{FIG1}, where a rectangular moving magnetic field pulse is applied to the permalloy strip with a magnitude of $10$ mT along the $+z$-direction. Our numerical results show that the interfacial DMI leads to a reduction of the threshold velocity of moving source, i.e. the minimum SW phase velocity, for coherent SW excitation in the absence of external ac magnetic field. Therefore it provides a promising route to reducing SCE threshold to facilitate experimental realization of such effect in magnetic medium.

\vbox{}
\section{Results}

\begin{figure}[t]
\centerline{\includegraphics[width=0.50\textwidth]{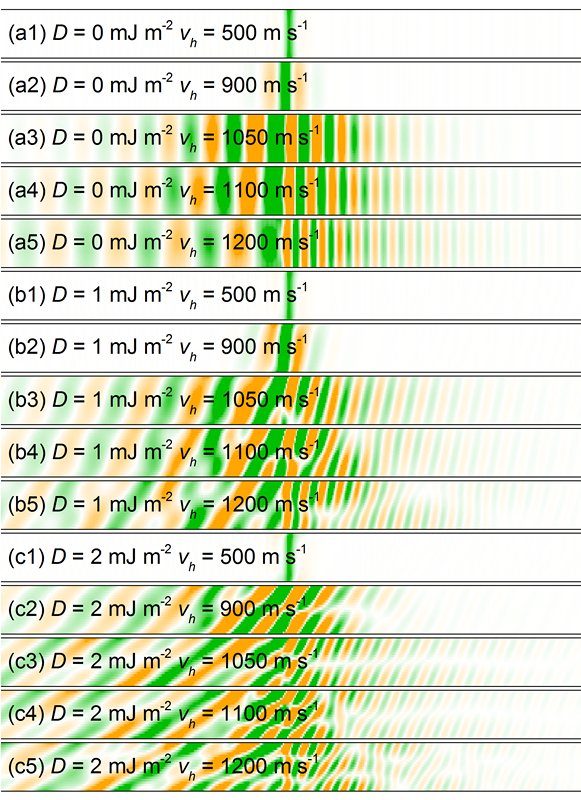}}
\caption{\textbf{Snapshots of the $z$-component of the magnetization in a magnetic strip in the vicinity of the field pulse traveling at a constant speed.}
(\textbf{a1}) $D=0$ mJ m$^{-2}$ and $v_{h}=500$ m s$^{-1}$,
(\textbf{a2}) $D=0$ mJ m$^{-2}$ and $v_{h}=900$ m s$^{-1}$,
(\textbf{a3}) $D=0$ mJ m$^{-2}$ and $v_{h}=1050$ m s$^{-1}$,
(\textbf{a4}) $D=0$ mJ m$^{-2}$ and $v_{h}=1100$ m s$^{-1}$,
(\textbf{a5}) $D=0$ mJ m$^{-2}$ and $v_{h}=1200$ m s$^{-1}$,
(\textbf{b1}) $D=1$ mJ m$^{-2}$ and $v_{h}=500$ m s$^{-1}$,
(\textbf{b2}) $D=1$ mJ m$^{-2}$ and $v_{h}=900$ m s$^{-1}$,
(\textbf{b3}) $D=1$ mJ m$^{-2}$ and $v_{h}=1050$ m s$^{-1}$,
(\textbf{b4}) $D=1$ mJ m$^{-2}$ and $v_{h}=1100$ m s$^{-1}$,
(\textbf{b5}) $D=1$ mJ m$^{-2}$ and $v_{h}=1200$ m s$^{-1}$,
(\textbf{c1}) $D=2$ mJ m$^{-2}$ and $v_{h}=500$ m s$^{-1}$,
(\textbf{c2}) $D=2$ mJ m$^{-2}$ and $v_{h}=900$ m s$^{-1}$,
(\textbf{c3}) $D=2$ mJ m$^{-2}$ and $v_{h}=1050$ m s$^{-1}$,
(\textbf{c4}) $D=2$ mJ m$^{-2}$ and $v_{h}=1100$ m s$^{-1}$,
(\textbf{c5}) $D=2$ mJ m$^{-2}$ and $v_{h}=1200$ m s$^{-1}$.
}
\label{FIG2}
\end{figure}

\noindent
\textbf{SW excitation via SCE in permalloy strips.}
The response of the magnetization distribution to the pulse velocity $v_h$ is shown in Fig.~\ref{FIG2}. A localized magnetic field of constant magnitude traveling along the wire axial direction with velocity of $v_h$, is applied to mimic the interacting force with the magnetization. The magnetization dynamics is strongly dependent on the pulse velocity $v_h$. For $D=0$ mJ m$^{-2}$ and $v_{h}=500$ m s$^{-1}$ and $900$ m s$^{-1}$ (see Fig.~\ref{FIG2}(a1-a2)), the moving magnetic field pulse only causes a distortion of the magnetization distribution traveling with the pulse. There is no SW excitation by the moving pulse. As the velocity of the magnetic pulse increases, SW excitations are observed for $v_{h}=1050$ m s$^{-1}$, $1100$ m s$^{-1}$, and $1200$ m s$^{-1}$, as shown in Fig.~\ref{FIG2}(a3-a5). At $v_{h}=1050$ m s$^{-1}$, the SWs proceeding the source and lagging the source are well-distinguished with different wave lengths. Upon the application of a moving dc magnetic field pulse, the system reaches a dynamic equilibrium and the excited spin waves comprise of two branches, giving rise to the SCE as reported in Ref.~\onlinecite{9}.

For the permalloy nanostrip with interfacial DMI, similar magnetization dynamics occur as the moving field pulse is applied. For the case of $D=1$ mJ m$^{-2}$, there is no SW excitation by the moving magnetic field with the speed of $v_{h}=500$ m s$^{-1}$ and $900$ m s$^{-1}$. The Cherenkov emission of SWs emerge at $v_{h}=1050$ m s$^{-1}$, $1100$ m s$^{-1}$, and $1200$ m s$^{-1}$ in the permalloy strip as shown in Fig.~\ref{FIG2}(b3-b5). The SWs excited in permalloy nanostrip are distorted due to the presence of DMI. The interfacial DMI term for any in-plane direction $\textbf{u}$ can be expressed as $D\textbf{\textit{z}}\times\textbf{u}$~\cite{28}, which may be treated as an effective field transverse to the magnetic track, resulting in the distorted spin waves. As $D=2$ mJ m$^{-2}$, spin wave is excited when $v_{h}=900$ m s$^{-1}$, indicating that the presence of interfacial DMI leads to the decreasing of the threshold velocity of the moving field source for spin wave excitation. It should be noticed that the SWs with more significant distortions are observed at $v_{h}=1050$ m s$^{-1}$, $1100$ m s$^{-1}$, and $1200$ m s$^{-1}$ as shown in Fig.~\ref{FIG2}(c3-c5). Moreover, the difference between the SW branches proceeding and lagging the source becomes even more obvious with increasing DMI strength for a given pulse velocity.

\begin{figure}[t]
\centerline{\includegraphics[width=0.50\textwidth]{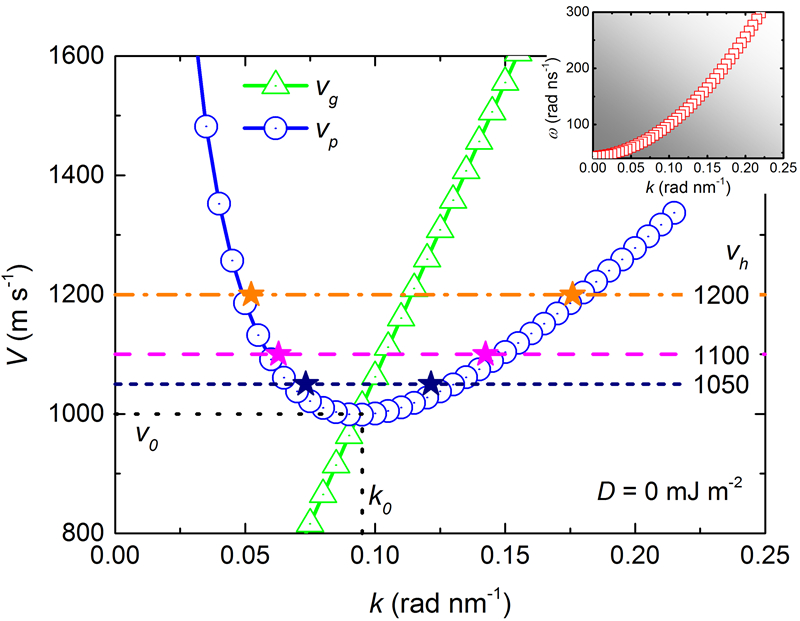}}
\caption{\textbf{Numerically determined phase velocity $v_{p}(k)$ and group velocity $v_{g}(k)$ of SWs in the magnetic strip which is $12$ $\mu$m long, $100$ nm wide, and $10$ nm thick as $D=0$ mJ m$^{-2}$.}
The value of $v_{p}(k)$ and $v_{g}(k)$ are extracted from the SW dispersion relation $\omega(k)$ shown in the inset. $v_{p}(k)$ has a minimum $v_0$ at $k_0$, where the two curves $v_{p}(k)$ and $v_{g}(k)$ cross. The colored stars represent the wave vectors of the SW tails excited by the moving field pulse applied to the strip at the corresponding speed. The colored horizontal line, indicating the speed of the field pulse, connect the two SW branches.
}
\label{FIG3}
\end{figure}

Figure~\ref{FIG3} shows the numerically determined SW phase velocity $v_{p}(k)$ and group velocity $v_{g}(k)$ for $D=0$ mJ m$^{-2}$. The inset of Fig.~\ref{FIG3} shows the SW dispersion in the permalloy strip, which agrees well with the analytical results for zero DMI (proportional to $k^{2}$). In Fig.~\ref{FIG3}, the phase velocity $v_{p}$ and group velocity $v_{g}$ of the SWs are extracted from the dispersion relation of $v_{p}(k)=\omega/k$ and $v_{g}(k)=d\omega/dk$. The minimum of the phase velocity occurs around $1000$ m s$^{-1}$, meaning that no SW is excited by the moving field pulse in the permalloy strip for the pulse velocities below the minimum $v_0$. In other words, the velocity $v_0$ is the threshold velocity of SCE, below which there is no coherent SWs excitation. As the field pulse moves at a velocity larger than $v_0$, there are two SW modes excited with different group velocity $v_{g}$ but equal phase velocity $v_{p}$. The SW packet with larger $\vec{k}$ ($v_{g}>v_{p}$) moves in the front of the source and leaves the one with smaller $\vec{k}$ ($v_{g}<v_{p}$) behind. The field pulses with different velocity $v_{h}$ are applied and the corresponding responses of the magnetization distribution are shown in Fig.~\ref{FIG2}(a1-a5). The calculated $k$ are shown with color stars in Fig.~\ref{FIG3}. As $v_{h}$ is below the critical velocity, there is no SW excitation by the moving field pulse. When $v_{h}=1050$ m s$^{-1}$, there are two SW modes observed in the permalloy strip. The spin waves formed in front of and behind the pulse exhibit different characteristics in Fig.~\ref{FIG2}(a3), and the corresponding $k$ values are shown in Fig.~\ref{FIG3}. The Cherenkov emission of SWs can also be observed for the case of $v_{h}=1100$ m s$^{-1}$ and $1200$ m s$^{-1}$. The dispersions are in an excellent agreement with the curves of $v_{p}(k)$ and $v_{g}(k)$. The curves of $v_{p}(k)$ and $v_{g}(k)$ extracted from the numerical SW dispersion relation $\omega(k)$, which is obtained by applying a localized ac field and extracting the wavelength of the excited spin waves, can predict the Cherenkov excitation of SWs precisely.

\begin{figure}[t]
\centerline{\includegraphics[width=0.50\textwidth]{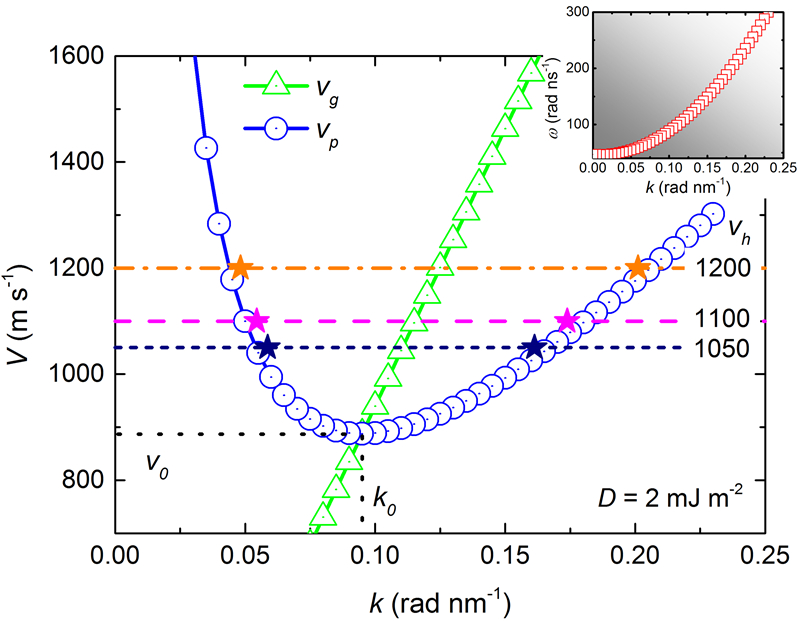}}
\caption{\textbf{Numerically determined phase velocity $v_{p}(k)$ and group velocity $v_{g}(k)$ of SWs in the magnetic strip, with a length of $12$ $\mu$m, a width of $100$ nm, and a thickness of $10$ nm when $D=2$ mJ m$^{-2}$.}
The value of $v_{p}(k)$ and $v_{g}(k)$ are extracted from the SW dispersion relation $\omega(k)$ shown in the inset. $v_{p}(k)$ has a minimum $v_0$ at $k_0$, where the curve $v_{p}(k)$ crosses the curve $v_{g}(k)$. The colored stars show the wave vectors of the SW tails excited by the moving field pulse applied to the strip at the corresponding speed. The colored horizontal line, indicating the speed of the field pulse, connects the two SW branches.
}
\label{FIG4}
\end{figure}

\vbox{}
\noindent
\textbf{Influence of DMI on SCE in permalloy.}
Figure~\ref{FIG4} shows the numerically determined phase velocity $v_{p}(k)$ and group velocity $v_{g}(k)$ with $D=2$ mJ m$^{-2}$, where $v_{p}(k)$ and $v_{g}(k)$ are calculated from the SW dispersion relation $\omega(k)$ shown in the inset. Similar to the case of $D=0$ mJ m$^{-2}$, there exists a critical velocity $v_0$ for the Cherenkov excitation of SWs. The Cherenkov excitation of SWs can be observed for $v_{h}>v_{0}$ ($887$ m s$^{-1}$). We also investigate the evolution of the SW branches by varying $v_{h}$ for $D=2$ mJ m$^{-2}$. At $v_{h}=1050$ m s$^{-1}$, the Cherenkov excitation of SWs is observed with two SW modes formed in the permalloy strip. The calculated $k$ marked with color stars agrees very well with the numerically determined curve of $v_{p}(k)$. The curves of $v_{p}(k)$ and $v_{g}(k)$ extracted from the SW dispersion relation $\omega(k)$ can predict the Cherenkov excitation of SWs precisely with the presence of finite DMI.

\begin{figure}[t]
\centerline{\includegraphics[width=0.50\textwidth]{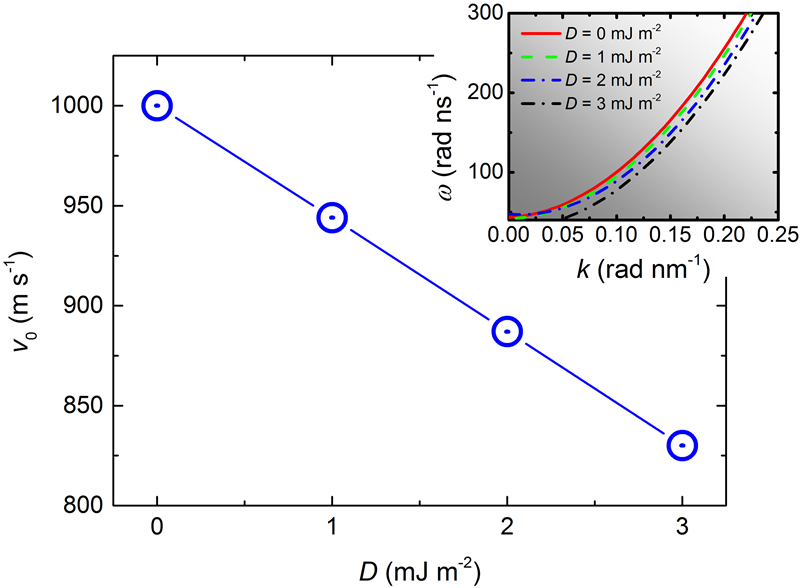}}
\caption{\textbf{The minimum velocity $v_{0}$ for the Spin-Cherenkov excitation as $D=0$ mJ m$^{-2}$, $1$ mJ m$^{-2}$, $2$ mJ m$^{-2}$, and $3$ mJ m$^{-2}$.}
The insets shows the corresponding dispersion relations, which are obtained by applying a localized ac magnetic field and extracting the wavelength of the excited spin waves.
}
\label{FIG5}
\end{figure}

Figure~\ref{FIG5} shows the minimum velocity $v_0$ of the Spin-Cherenkov excitation for $D=0$ mJ m$^{-2}$, $1$ mJ m$^{-2}$, $2$ mJ m$^{-2}$ and $3$ mJ m$^{-2}$, as well as the corresponding dispersion relation. Only the branches of positive $k$ are shown. The critical velocity for the SCE decreases linearly with increasing DMI in the investigated DMI range. As shown in the inset of Fig.~\ref{FIG5}, the dispersion relation for $D=1$ mJ m$^{-2}$ deviates from the case of $D=0$ mJ m$^{-2}$, resulting in a smaller critical velocity $v_0$. $v_{0}=1000$ m s$^{-1}$ for $D=0$ mJ m$^{-2}$ whereas $v_{0}=944$ m s$^{-1}$ for $D=1$ mJ m$^{-2}$. The critical velocity further drops to $887$ m s$^{-1}$ as $D$ increases to $2$ mJ m$^{-2}$, indicating that the Cherenkov emission of SWs in permalloy strip with finite DMI can be excited more easily than the case of zero DMI. The SW dispersion relations without DMI ($D=0$ mJ m$^{-2}$) is a parabolic function of SW vector $k$ when $k$ is large~\cite{9,13,29,30}. In Refs.~\onlinecite{14,25}, the SW dispersion relation is given analytically when DMI is included, indicating that the interfacial DMI results in the asymmetric dispersion. Such asymmetric dependence due to the interfacial DMI depends on the spin wave vector $k$ and equilibrium magnetization distributions. Different with the typical Damon-Eshbach spin waves studied in Refs.~\onlinecite{14,25}, the SWs in our case propagate along the $\pm x$-directions while the equilibrium magnetization is along the $+x$-direction, corresponding to the back-volume mode. The SW dispersion is given by $\omega=\gamma_{0}\sqrt{\omega_{0}^{2}-(kD^{*}m_{z0})^{2}}$ when the DMI is considered. Here $D^{*}=\frac{2D}{\mu_{0}M_{\text{S}}}$ and $\omega_{0}$ is the angular frequency in the absence of DMI. $m_{z0}$ is the $z$-component of the magnetization. The SW dispersion relation remains symmetric, when the interfacial DMI is included. The SCE threshold can be calculated by $\frac{d\omega}{dk}=\frac{\omega}{k}$, which decreases with DMI~\cite{9}.

\vbox{}
\section{Discussion}

\noindent
The Cherenkov emission of spin waves has been numerically studied by considering the effect of DMI in permalloy strip.
The resonant Spin-Cherenkov effect can be excited as the velocity of moving magnetic field pulse exceeds a certain threshold velocity.
The Spin-Cherenkov effect threshold can be reduced in the presence of finite DMI.
By further tuning the material parameters and geometries, resonant spin waves can be excited through Spin-Cherenkov effect at much lower threshold velocity of the moving dc field source.
\blue{On the other hand, it is also feasible to increase the interaction region of the moving dc field source on the magnetic strip in real experiments.}
\blue{Indeed, we have investigated the Spin-Cherenkov effect with different sizes of the moving dc field source (see Supplementary Figure 1), as well as with different thicknesses of the magnetic strip (see Supplementary Figure 2).}
The Spin-Cherenkov effect in finite-DMI system might be interesting for fundamental physics and also promising for potential applications in spintronic and magnonic devices, easing the experimental complexity and difficulty of applying an external ac magnetic field or current for resonant SW excitations.

\vbox{}
\section{Methods}

\noindent
\textbf{Modeling and simulation.}
The micromagnetic simulations are performed using the Object Oriented MicroMagnetic Framework (OOMMF) software including the interface-induced Dzyaloshinskii-Moriya interaction (DMI) extension module~\cite{19,20,21,31,32,33}.
The three-dimensional time-dependent magnetization dynamics is controlled by the Landau-Lifshitz-Gilbert (LLG) ordinary differential equation~\cite{34,35,36}
\begin{equation}
\frac{d\mathbf{M}}{dt}=-\gamma_{\text{0}}\mathbf{M}\times\mathbf{H}_{\text{eff}}+\frac{\alpha}{M_{\text{S}}}(\mathbf{M}\times\frac{d\mathbf{M}}{dt}),
\label{LLG}
\end{equation}
where $\textbf{M}$ is the magnetization, $\textbf{H}_{\text{eff}}$ is the effective field, $\gamma_0$ is the Gilbert gyromagnetic ration, and $\alpha$ is the Gilbert damping coefficient.
The effective field reads as follows
\begin{equation}
\mathbf{H}_{\text{eff}}=-\mu_{0}^{-1}\frac{\partial E}{\partial \mathbf{M}}.
\label{EffectiveField}
\end{equation}
The average energy density $E$ as a function of $\textbf{M}$ is given by
\begin{align}
E=A[\nabla(\frac{\mathbf{M}}{M_{\text{S}}})]^{2}&-K\frac{(\mathbf{n}\cdot\mathbf{M})^{2}}{M_{\text{S}}^{2}}-\mu_{0}\mathbf{M}\cdot\mathbf{H} \notag \\
&-\frac{\mu_{0}}{2}\mathbf{M}\cdot\mathbf{H}_{\text{d}}(\mathbf{M})+E_{\text{DM}},
\label{EnergyDensity}
\end{align}
where $A$ and $K$ are the exchange and anisotropy energy constants, respectively.
$\textbf{H}$ and $\textbf{H}_{\text{d}}(\textbf{M})$ are the applied and magnetostatic self-interaction fields while $M_{\text{S}}=|\textbf{M}(\text{\textbf{r}})|$ is the spontaneous magnetization.
The $E_{\text{DM}}$ is the energy density of the interfacial DMI of the form~\cite{21,31,36,37}
\begin{align}
E_{\text{DM}}=\frac{D}{M_{\text{S}}^{2}}(&M_{z}\frac{\partial M_{x}}{\partial x}+M_{z}\frac{\partial M_{y}}{\partial y} \notag \\
-&M_{x}\frac{\partial M_{z}}{\partial x}-M_{y}\frac{\partial M_{z}}{\partial y}),
\label{DMIDensity}
\end{align}
where the $M_x$, $M_y$, $M_z$ are the components of the magnetization $\textbf{M}$ and $D$ is the interfacial DMI constant.
The five terms at the right side of equation~(\ref{EnergyDensity}) correspond to the exchange energy, the anisotropy energy, the applied field (Zeeman) energy, the magnetostatic (demagnetization) energy and the interfacial DMI energy, respectively.

The typical material parameters of permalloy, $\mu_{0}M_{\text{S}}=1$ T, exchange constant $A=1.3\times 10^{-11}$ J m$^{-1}$, and zero anisotropy are adopted~\cite{8,9,22,26}.
Considering that the value of the effective interfacial DMI constant in the Py/Pt bilayers has been estimated to be within the range of $1.0\sim 2.2$ mJ m$^{-2}$ in Refs.~\onlinecite{27,38}, the interfacial DMI constant $D$ is varied from $0$ to $3$ mJ m$^{-2}$ in this paper.
A rectangular shape field pulse is applied to $12$-$\mu$m-long, $100$-nm-wide, and $10$-nm-thick magnetic strip with a magnitude of $10$ mT in the $+z$-direction and a $12$ nm width in the $x$-direction, as shown in Fig.~\ref{FIG1}.
The moving field pulse can be realized, for example, with a laser beam scanning over the surface of magnetic thin films~\cite{39}.
\blue{The results with different widths of the moving field pulse are shown in Supplementary Figure~1.}
\blue{The results with different thicknesses of the magnetic strip are shown in Supplementary Figure~2.}
For simplicity, the uniform field along the film thickness is assumed.
All samples are discretized into cells of $3$ nm $\times$ $5$ nm $\times$ $5$ nm in the simulation.
Gilbert damping coefficient $\alpha$ is set to be $0.02$ and the value for Gilbert gyromagnetic ratio $\gamma_0$ equals $2.211\times 10^{5}$ m A$^{-1}$ s$^{-1}$.
Initially, the magnetization orients along the $+x$-direction due to the shape anisotropy.
The absorbing boundary condition has been implemented at both ends of the nanostrip, which effectively avoids the spurious spin wave reflections~\cite{40}.



\section{Acknowledgements}
\noindent
X.Z. was supported by JSPS RONPAKU (Dissertation Ph.D.) Program and was partially supported by the Scientific Research Fund of Sichuan Provincial Education Department (Grant No.~16ZA0372).
M.Y. acknowledges the support by National Natural Science Foundation of China (Project No.~11374203).
Y.Z. acknowledges the support by National Natural Science Foundation of China (Project No.~1157040329), the Seed Funding Program for Basic Research and Seed Funding Program for Applied Research from the HKU, ITF Tier 3 funding (ITS/171/13 and ITS/203/14), the RGC-GRF under Grant HKU 17210014, and University Grants Committee of Hong Kong (Contract No.~AoE/P-04/08).
W.S.Z. acknowledges the support by the projects from the Chinese Postdoctoral Science Foundation (No.~2015M570024), National Natural Science Foundation of China (Projects No.~61501013, No.~61471015 and No.~61571023), Beijing Municipal Commission of Science and Technology (Grant No.~D15110300320000), and the International Collaboration Project (No.~2015DFE12880) from the Ministry of Science and Technology of China.
X.Z. thanks M.J. Donahue for useful discussions.

\section{Author Contributions}
\noindent
Y.Z. conceived the problem. Y.Z. and W.S.Z. coordinated the project. X.Z. carried out the numerical simulations. J.X. performed the theoretical analysis with the input from M.Y. All authors discussed the results and prepared the manuscript.

\section{Additional Information}
\noindent
Supplementary information accompanies this paper at http://www.nature.com/srep

\vbox{}\noindent
Correspondence and requests for materials should be addressed to W.Z. and Y.Z.

\section{Competing Financial Interests}
\noindent
The authors declare no competing financial interests.


\begin{thebibliography}{99}
\bibitem{1} \v{C}erenkov, P. A. Visible radiation produced by electrons moving in a medium with velocities exceeding that of light. \textit{Phys. Rev.} \textbf{52}, 378-379 (1937).
\bibitem{2} Jelley, J. V. Cerenkov radiation and its applications. \textit{Br. J. Appl. Phys.} \textbf{6}, 227-232 (1955).
\bibitem{3} Berger, H. Complex Doppler effect in dispersive media. \textit{Am. J. Phys.} \textbf{44}, 851-854 (1976).
\bibitem{4} Hu, X., Hang, Z., Li, J., Zi, J. \& Chan, C. T. Anomalous Doppler effects in phononic band gaps. \textit{Phys. Rev. E} \textbf{73}, 015602 (2006).
\bibitem{5} Lisenkov, I. V. \& Nikitov, S. A. The complex Doppler effect in double negative media. \textit{J. Commun. Technol. Electron.} \textbf{56}, 687-689 (2011).
\bibitem{6} Vlaminck, V. \& Bailleul, M. Current-induced spin-wave Doppler Shift. \textit{Science} \textbf{322}, 410-413 (2008).
\bibitem{7} Sekiguchi, K. \textit{et al.} Time-domain measurement of current-induced spin wave dynamics. \textit{Phys. Rev. Lett.} \textbf{108}, 017203 (2012).
\bibitem{8} Yan, M., Andreas, C., K\'{a}kay, A., Garc\'{i}a-S\'{a}nchez, F. \& Hertel, R. Fast domain wall dynamics in magnetic nanotubes: suppression of Walker breakdown and Cherenkov-like spin wave emission. \textit{Appl. Phys. Lett.} \textbf{99}, 122505 (2011).
\bibitem{9} Yan, M., K\'{a}kay, A., Andreas, C. \& Hertel, R. Spin-Cherenkov effect and magnonic mach cones. \textit{Phys. Rev. B} \textbf{88}, 220412 (2013).
\bibitem{10} Zakeri, K. \textit{et al.} Asymmetric spin-wave dispersion on Fe(110): direct evidence of the Dzyaloshinskii-Moriya interaction. \textit{Phys. Rev. Lett.} \textbf{104}, 137203 (2010).
\bibitem{11} Nagaosa, N. \& Tokura, Y. Topological properties and dynamics of magnetic skyrmions. \textit{Nat. Nanotech.} \textbf{8}, 899-911 (2013).
\bibitem{12} Fert, A., Cros, V. \& Sampaio, J. Skyrmions on the track. \textit{Nat. Nanotech.} \textbf{8}, 152-156 (2013).
\bibitem{13} Cortes-Ortuno, D. \& Landeros, P. Influence of the Dzyaloshinskii-Moriya interaction on the spin-wave spectra of thin films. \textit{J. Phys.: Condens. Matter.} \textbf{25}, 156001 (2013).
\bibitem{14} Moon, J.-H. \textit{et al.} Spin-wave propagation in the presence of interfacial Dzyaloshinskii-Moriya interaction. \textit{Phys. Rev. B} \textbf{88}, 184404 (2013).
\bibitem{15} Cho, J. \textit{et al.} Thickness dependence of the interfacial Dzyaloshinskii-Moriya interaction in inversion symmetry broken systems. \textit{Nat. Commun.} \textbf{6}, 7635 (2015).
\bibitem{16} Kravchuk, V. P. Influence of Dzialoshinskii-Moriya interaction on static and dynamic properties of a transverse domain wall. \textit{J. Magn. Magn. Mater.} \textbf{367}, 9-14 (2014).
\bibitem{17} Wang, W. \textit{et al.} Magnon-driven domain-wall motion with the Dzyaloshinskii-Moriya interaction. \textit{Phys. Rev. Lett.} \textbf{114}, 087203 (2015).
\bibitem{18} Zhou, Y. \& Ezawa, M. A reversible conversion between a skyrmion and a domain-wall pair in junction geometry. \textit{Nat. Commun.} \textbf{5}, 4652 (2014).
\bibitem{19} Zhang, X., Ezawa, M. \& Zhou, Y. Magnetic skyrmion logic gates: conversion, duplication and merging of skyrmions. \textit{Sci. Rep.} \textbf{5}, 9400 (2015).
\bibitem{20} Zhang, X. \textit{et al.} All-magnetic control of skyrmions in nanowires by a spin wave. \textit{Nanotechnology} \textbf{26}, 225701 (2015).
\bibitem{21} Rohart, S. \& Thiaville, A. Skyrmion confinement in ultrathin film nanostructures in the presence of Dzyaloshinskii-Moriya interaction. \textit{Phys. Rev. B} \textbf{88}, 184422 (2013).
\bibitem{22} Im, M. Y. \textit{et al.} Symmetry breaking in the formation of magnetic vortex states in a permalloy nanodisk. \textit{Nat. Commun.} \textbf{3}, 983 (2012).
\bibitem{23} M\"{u}hlbauer, S. \textit{et al.} Skyrmion lattice in a chiral magnet. \textit{Science} \textbf{323}, 915-919 (2009).
\bibitem{24} Huang, S. X. \& Chien, C. L. Extended skyrmion phase in epitaxial FeGe(111) thin films. \textit{Phys. Rev. Lett.} \textbf{108}, 267201 (2012).
\bibitem{25} Di, K. \textit{et al.} Direct observation of the Dzyaloshinskii-Moriya interaction in a Pt/Co/Ni film. \textit{Phys. Rev. Lett.} \textbf{114}, 047201 (2015).
\bibitem{26} Chen, S. J. \textit{et al.} Effect of Dzyaloshinskii-Moriya interaction on the magnetic vortex oscillator driven by spin-polarized current. \textit{J. Appl. Phys.} \textbf{117}, 17B720 (2015).
\bibitem{27} Stashkevich, A. A. \textit{et al.} Experimental study of spin-wave dispersion in Py/Pt film structures in the presence of an interface Dzyaloshinskii-Moriya interaction. \textit{Phys. Rev. B} \textbf{91}, 214409 (2015).
\bibitem{28} Boulle, O. \textit{et al.} Domain wall tilting in the presence of the Dzyaloshinskii-Moriya interaction in out-of-plane magnetized magnetic nanotracks. \textit{Phys. Rev. Lett.} \textbf{111}, 217203 (2013).
\bibitem{29} You, C.-Y. \& Kim, N.-H. Critical Dzyaloshinskii-Moriya interaction energy density for the skyrmion states formation in ultrathin ferromagnetic layer. \textit{Curr. Appl. Phys.} \textbf{15}, 298-301 (2015).
\bibitem{30} You, C.-Y. Curie temperature of ultrathin ferromagnetic layer with Dzyaloshinskii-Moriya interaction. \textit{J. Appl. Phys.} \textbf{116}, 053902 (2014).
\bibitem{31} Sampaio, J., Cros, V., Rohart, S., Thiaville, A. \& Fert, A. Nucleation, stability and current-induced motion of isolated magnetic skyrmions in nanostructures. \textit{Nat. Nanotech.} \textbf{8}, 839-844 (2013).
\bibitem{32} Donahue, M. J. \& Porter, D. G. \textit{OOMMF User's Guide, Version 1.0 Interagency Report NISTIR 6376} (National Institute of Standards and Technology, Gaithersburg, MD, 1999).
\bibitem{33} Boulle, O., Buda-Prejbeanu, L. D., Ju\'{e}, E., Miron, I. M. \& Gaudin, G. Current induced domain wall dynamics in the presence of spin orbit torques. \textit{J. Appl. Phys.} \textbf{115}, 17D502 (2014).
\bibitem{34} Gilbert, T. L. A Lagrangian formulation of the gyromagnetic equation of the magnetization field. \textit{Phys. Rev.} \textbf{100}, 1243 (1955).
\bibitem{35} Landau, L. \& Lifshitz, E. On the theory of the dispersion of magnetic permeability in ferromagnetic bodies. \textit{Physik. Z. Sowjetunion} \textbf{8}, 153-169 (1935).
\bibitem{36} Thiaville, A., Rohart, S., Ju\'{e}, \'{E}., Cros, V. \& Fert, A. Dynamics of Dzyaloshinskii domain walls in ultrathin magnetic films. \textit{Europhys. Lett.} \textbf{100}, 57002 (2012).
\bibitem{37} Bogdanov, A. N. \& Yablonskii, D. A. Thermodynamically stable "vortices" in magnetically ordered crystals. The mixed state of magnets. \textit{Zh. Eksp. Teor. Fiz.} \textbf{95}, 178-182 (1989).
\bibitem{38} Nembach, H. T., Shaw, J. M., Weiler, M., Ju\'{e}, E. \& Silva, T. J. Linear relation between Heisenberg exchange and interfacial Dzyaloshinskii-Moriya interaction in metal films. \textit{Nat. Phys.} \textbf{11}, 825-829 (2015).
\bibitem{39} Vorobev, P. V. \& Kolokolov, I. V. Cherenkov emission of magnons by a slow monopole. \textit{JETP Lett.} \textbf{67}, 910-912 (1998).
\bibitem{40} Venkat, G., Franchin, M., Fangohr, H. \& Prabhakar, A. Mesh size and damped edge effects in micromagnetic spin wave simulations. \textit{arXiv} \textbf{1405.4615}, http://arxiv.org/abs/1405.4615 (2014) (Accessed: 1st June 2014).
\end{thebibliography}
\end{document}